\documentclass{WileyMSP-template}
\usepackage{amsfonts}
\usepackage{mathrsfs}
\usepackage{amsmath}
\usepackage{color}
\usepackage{graphicx}
\usepackage{bm}
\usepackage{amssymb}
\usepackage{xspace}
\usepackage{epstopdf}
\usepackage{dcolumn}
\usepackage{multirow}
\usepackage{wrapfig}
\usepackage{cite}
\usepackage{ragged2e}
\usepackage{makecell}

\begin{document}


\title{3D carbon allotropes: Topological quantum materials with obstructed atomic insulating phases, multiple bulk-boundary correspondences, and real topology}

\maketitle


\author{Jianhua Wang$^{\#}$,}
\author{Ting-Ting Zhang$^{\#}$,}
\author{Qianwen Zhang,}
\author{Xia Cheng,}
\author{Wenhong Wang,}
\author{Shifeng Qian*,}
\author{Zhenxiang Cheng*,}
\author{Gang Zhang,}
\author{Xiaotian Wang*}\\

($^{\#}$These authors contributed equally to this work.)


\dedication{ }

\begin{affiliations}
J. Wang\\
School of Material Science and Engineering, Tiangong University, Tianjin 300387, China\\
Institute for Superconducting and Electronic Materials (ISEM), University of Wollongong, Wollongong 2500, Australia\\
T. Zhang\\
Beijing National Laboratory for Condensed Matter Physics, Institute of Physics, Chinese Academy of Sciences, Beijing 100190, China\\
Q. Zhang, X. Cheng\\
School of Physical Science and Technology, Southwest University, Chongqing 400715, China\\
W. Wang\\
School of Material Science and Engineering, Tiangong University, Tianjin 300387, China\\
S. Qian\\
Anhui Province Key Laboratory of Optoelectric Materials Science and Technology, Department of Physics, Anhui Normal University, Anhui, Wuhu 241000, China\\
Email:qiansf@ahnu.edu.cn\\
Z. Cheng\\
Institute for Superconducting and Electronic Materials (ISEM), University of Wollongong, Wollongong 2500, Australia\\
Email:cheng@uow.edu.au\\
G. Zhang\\
Institute of High-Performance Computing, Agency for Science, Technology and Research (A$\*$STAR), 138632 Singapore, Singapore\\
X. Wang\\
Institute for Superconducting and Electronic Materials (ISEM), University of Wollongong, Wollongong 2500, Australia\\
School of Physical Science and Technology, Southwest University, Chongqing 400715, China\\
Email:xiaotianw@uow.edu.au\\
\end{affiliations}



\begin{abstract}
The study of topological phases with unconventional bulk-boundary correspondences and nontrivial real Chern number has garnered significant attention in the topological states of matter. Using the first-principle calculations and theoretical analysis, we perform a high-throughput material screening of the 3D obstructed atomic insulators (OAIs) and 3D real Chern insulators (RCIs) based on the Samara Carbon Allotrope Database (SACADA). Results show that 422 out of 703 3D carbon allotropes are 3D OAIs with multiple bulk-boundary correspondences, including 2D obstructed surface states (OSSs) and 1D hinge states, which are in one dimension and two dimensions lower than the 3D bulk, respectively. The 2D OSSs in these OAIs can be modified when subjected to appropriate boundaries, which benefits the investigation of surface engineering and  the development of efficient topological catalysts. These 422 OAIs, which have 2D and 1D boundary states, are excellent platforms for multi-dimensional topological boundaries research. Remarkably, 138 of 422 OAIs are also 3D RCIs, which show a nontrivial real topology in the protection of spacetime inversion symmetry. Our work not only provides a comprehensive list of 3D carbon-based OAIs and RCIs, but also guides their application in various aspects based on multiple bulk-boundary correspondences and real topological phases.

\end{abstract}

\justifying

\section{Introduction}
In condensed-matter physics, band topology~\cite{add1,add2,add3,add4,add5} has flourished and developed over the last ten years. Many materials have been predicted as topological insulators~\cite{add6,add7,add8,add9,add10} or topological semimetals~\cite{add11,add12,add13,add14,add15,add16,add17,add18,add19,add20} with topological boundary states. Meanwhile, topological trivial insulators have lately received attention due to the advancement of topological quantum chemistry theory~\cite{add21,add22}. A topological trivial insulator is a material whose band representations (BRs) of valence bands can be characterized by a sum of elementary BRs (EBRs)~\cite{add21}, which means the trivial state can be described by a set of exponentially localized Wannier functions. Based on the Wyckoff positions where the orbitals responsible for inducing the band representations (BRs) are situated, topological trivial insulators can be classified into two categories: obstructed atomic insulators (OAIs) and atomic insulators~\cite{add23,add24,add25,add26,add27,add27a,add28,add30}. For the atomic insulators, their BRs are all induced from the orbitals that are located at atomic-occupied Wyckoff positions (WPs); however, for the OAIs, their BRs come from the orbitals that are located at atom-occupied and atom-unoccupied WPs~\cite{add31,add32}. Such atom-unoccupied WPs in OAIs are named obstructed Wannier charge centers (OWCCs)~\cite{add31,add32}. When the cleavage termination cuts through those OWCCs in a 3D OAI, the 2D obstructed surface states (OSSs) will occur~\cite{add27,add27a,add28}. Unlike topological insulators, their topological surface states are connected to the bulk states, which generally host small spin-orbit coupling-induced gaps~\cite{add6,add7}; for OAIs, their OSSs are usually clean, located between the large energy gap, and simply to distinguish in the energy spectrum from the bulk states, which is advantageous for theoretical analysis and experimental detection~\cite{add30}. Remarkably, it has been suggested that the 2D OSSs in a 3D OAI can be used as a descriptor to define catalytic active sites in 3D inorganic crystalline materials~\cite{add27}. That is,  the crystal surface with OSSs could be active for molecule bonding and adsorption. Moreover, the existence of OSSs in SrIn$_2$P$_2$~\cite{add30} has been validated by the utilization of scanning tunneling microscopy and angle-resolved photoemission spectroscopy.

  \begin{figure}[h]
\centering
  \includegraphics[height=7cm]{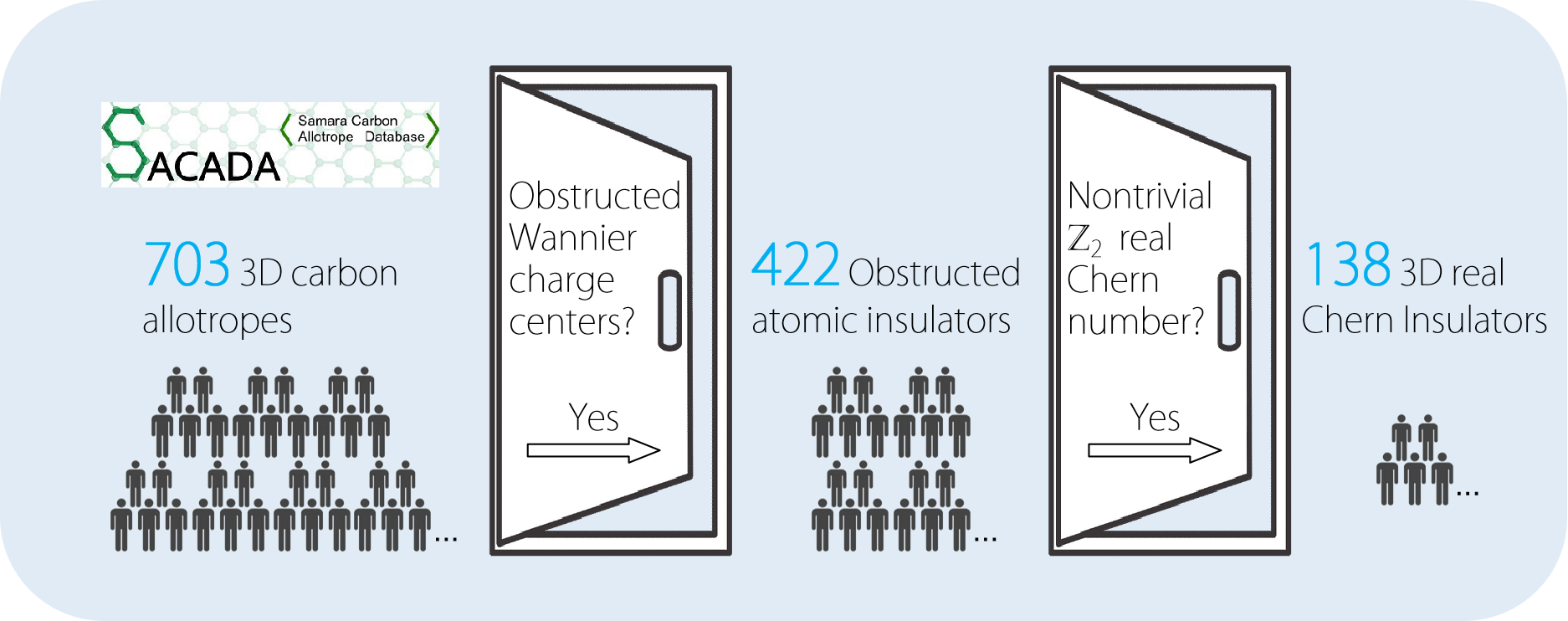}
  \caption{The schematic procedure for discovering the obstructed atomic insulators and real Chern insulators among 703 3D carbon allotropes based on the Samara Carbon Allotrope Database (SACADA).}
  \label{fig1}
\end{figure}

Note that the study of topological matters has recently revolutionized with the finding of second-order topological phases ~\cite{add33,add34,add35,add36,add37,add38,add39,add40,add41,add42,add43,add44,add45,add46,add47,add48,add49,add50}. Second-order topological phases in 3D induce unconventional bulk-boundary correspondence because they exhibit boundary states in two dimensions lower than the 3D bulk, as opposed to the previous first-order topological phases, which contained topological boundary states in one dimension lower than the 3D bulk. The second-order topological phases in 3D are manifested in the hinge modes located at the hinges of the sample materials with a tube geometry. So far, second-order topological phases in 3D have only been experimentally verified in Bi, Bi$_x$Sb$_{1-x}$, and Bi$_2$TeI  ~\cite{add34,add85, add50a}, and the scarcity of suitable materials hinders further investigations. For a 3D OAI, it is also interesting to study the existence of (3-2)D boundaries, i.e., 1D hinges, besides its (3-1)D boundaries, i.e., 2D OSSs.

Moreover, as guaranteed by certain symmetry constraints, such as the spacetime inversion symmetry ($\mathcal{P} \mathcal{T}$) and the absence of spin-orbit coupling (SOC) effect~\cite{add51,add52,add53}, the band eigenstates are required to be real ~\cite{add53,add54,add55,add56}. The real Chern insulator state (RCI) state, characterized by the nontrivial $\mathbb{Z}_2$ valued real Chern number $\nu_R$ (also known as the second Stiefel-Whitney number)~\cite{add54,add55}, has been revealed in a few two-dimensional systems~\cite{add57}. Notably, the concept can be extended to three dimensions. In this work, we term the 3D OAIs with a nontrivial real topology as 3D RCIs, where every 2D horizontal or vertical slice has a nontrivial $\nu_R$.

From a materials science perspective, finding ideal material candidates for 3D OAIs and 3D RCIs with multiple dimensions of boundaries and real topological phases continues to be a crucial subject. In this study, we aim to make advancements in predicting the properties of 3D OAIs and 3D RCIs in all forms of carbon allotropes by high-throughput material screening. In addition to the widely recognized carbon allotropes such as graphite~\cite{add58,add59,add60}, diamond~\cite{add61,add62}, carbon nanotubes~\cite{add63,add64,add65}, graphene~\cite{add66,add67,add68}, and fullerenes~\cite{add69,add70}, there have been over 700 more carbon allotropes that have been either theoretically predicted or experimentally synthesized. Through comprehensive examination utilizing first-principle calculations and theoretical analysis based on the Samara Carbon Allotrope Database (SACADA)~\cite{add71}, we have investigated a total of 703 carbon allotropes. Our findings indicate that among these candidates, 422 candidates are 3D OAIs.   In addition, out of these OAIs, 138 can be categorized as 3D $\mathcal{P} \mathcal{T}$-symmetric RCIs that possess a nontrivial $\nu_R$. Note that there have been no papers reporting the comprehensive analysis of all OAIs and RCIs for all forms of carbon allotropes to yet. Using 37-SG. 194-lon (No. 37 in SACADA and  space group (SG.) 194) and 35-dia-a (No. 35 in SACADA and SG. 224) as examples, we will show their OAI state and RCI state, respectively. Moreover, we shall examine the OSS states in the example of 37-SG. 194-lon by choosing four surface planes, i.e., (001)$_1$, $(10 \overline{1} 0)$, (001)$_2$, and (010) planes, as cleavage terminations, and find there would exist four completely different 2D OSSs on these surface planes. In other words, the 2D OSSs in 3D OAIs can be modified when subjected to appropriate boundaries, making them more ideal for on-demand surface engineering. Besides the 2D OSSs, the 1D hinge states on vertical hinges can also be found in 37-SG. 194-lon. That is, 2D and 1D boundaries (surface and hinge) coexisted in a single 37-SG. 194-lon, which provides a platform for studying boundaries across multiple dimensions. Unlike 37-SG. 194-lon, the 35-SG. 224-dia-a hosts a real Chern insulator state, characterized by a $\nu_R$, which can be viewed as a promising platform for exploring the fascinating physics of real topological phases.

\section{Results and discussion}

\subsection{Database screening scheme}

The crystallographic data for all carbon allotropes can be acquired from the SACADA database~\cite{add71}, which provides a comprehensive list of 703 3-periodic carbon allotropes.

The first step is identifying the 3D OAIs from the 703 3D carbon allotropes using the theory of elementary band representation (EBR) ~\cite{add21,add22}. The EBRs correspond to the smallest sets of band structures that can be derived from maximal localized atomic-like Wannier functions. A sum of EBRs can fully describe the band structures below the Fermi level of atomic insulators and OAIs. However, the OAI is the insulator with the EBRs at the WPs that are unoccupied by atoms.  Such atom-unoccupied WPs are known as OWCCs.

To see if the OWCCs are present in the 703 3D carbon allotropes, we estimated the BRs of occupied bands at high-symmetry points for each 703 carbon allotropes (see $\textbf{Figure~\ref{fig1}}$). To characterize the symmetry properties, the concept of symmetry-data vector $B$ is defined~\cite{add21}, and its explicit form is provided below:

\begin{equation}
B=\left\{n_1^1 G_{K_1}^1 \oplus n_2^2 G_{K_1}^2 \oplus \ldots, n_2^1 G_{K_2}^1 \oplus n_2^2 G_{K_2}^2 \oplus \cdots\right\},
\end{equation}
the $G_{K_i}^j$ is the $j$th irrep of the little group at the maximal momenta K$_i$, and  $n_i^j$ is the multiplicity of irrep $G_{K_i}^j$.

\begin{figure}[h]
\centering
  \includegraphics[height=10cm]{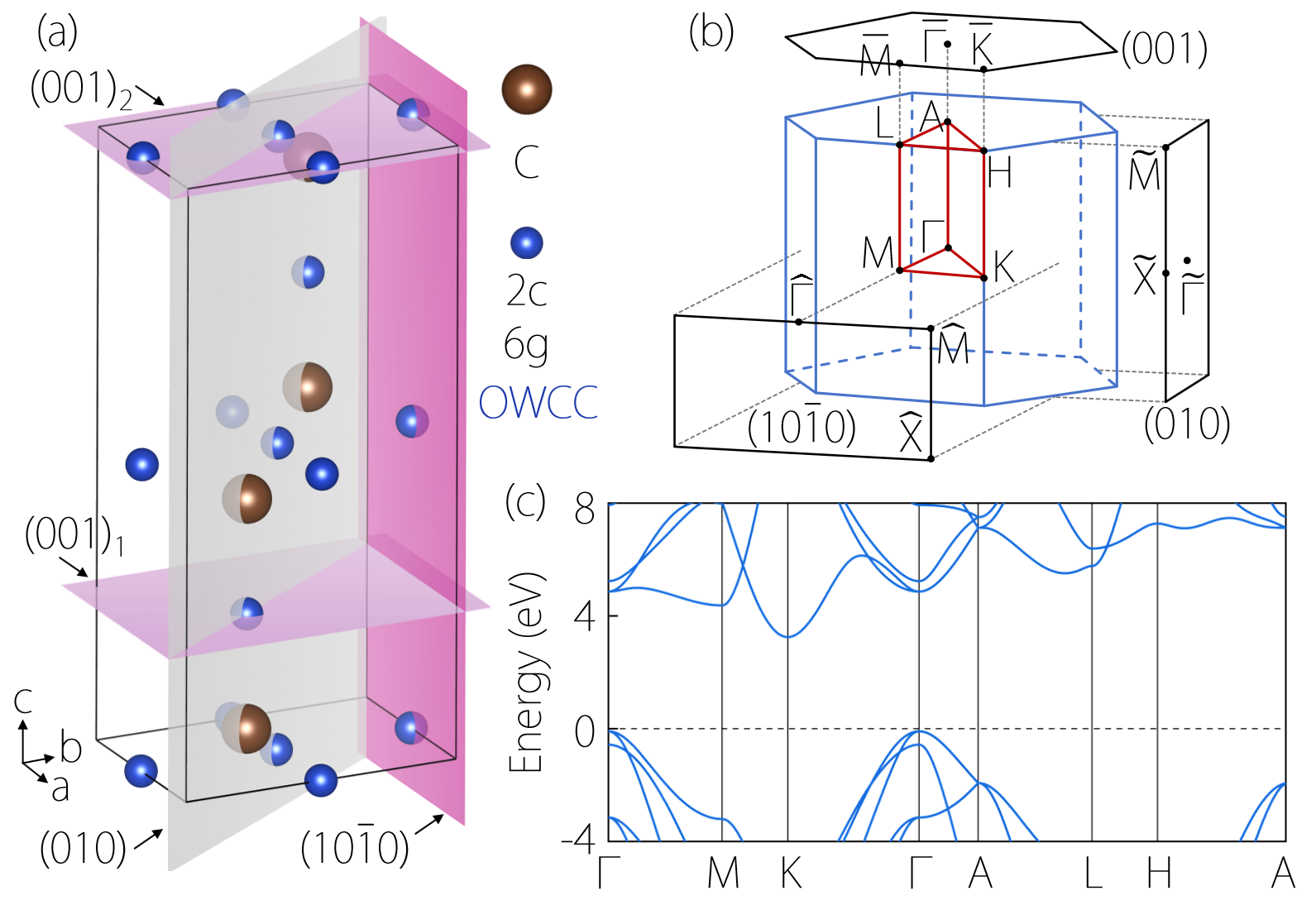}
  \caption{(a)  Crystal structure of 37-SG. 194-lon. In (a), the OWCCs residing at the Wyckoff positions 2c and 6g are denoted by small blue balls. Four surface terminations, including (001)$_1$, $(10 \overline{1} 0)$, (001)$_2$, and (010), are highlighted by arrows. (b) The 3D Brillouin zone (BZ) and the projected (001), $(10 \overline{1} 0)$, and (010) surfaces. (c) Calculated band structure for 37-SG. 194-lon along the symmetry paths $\Gamma$--M--K--$\Gamma$--A--L--H--A.}
  \label{fig2}
\end{figure}

$B$ can be decomposed on the basis of EBRs. The decomposition is subjected to

\begin{equation}
B=\sum_i p_i(E B R)_i.
\end{equation}
The coefficient $p_i$ can show the topological features of the system~\cite{add21}. For clarity, we select 37-SG. 194-lon, as an example, the $B$  and corresponding EBRs  for 37-SG. 194-lon are shown in $\textbf{Table 1}$. It is obvious that the coefficients $p_{\left(A_1^{\prime}\right)_{2 c}} \uparrow G=1$ and  $p_{\left(A_{g}\right)_{6 g}} \uparrow G=1$, and the others are zero, indicating the bands below the Fermi level can be expressed as a combination of EBRs: ${\left(A_1^{\prime}\right)_{2 c}} \uparrow G$ + $ {\left(A_{g}\right)_{6 g}} \uparrow G$. Hence, the OWCCs of  37-SG. 194-lon reside at the Wyckoff positions 2c and 6g, which are mismatching with atomic-occupied Wyckoff positions 4f (see $\textbf{Figure~\ref{fig2}}$a), reflecting 37-SG. 194-lon is an OAI. After screening, we found 422 out of 703 carbon allotropes exhibit the OWCCs, indicating the nature of the OAI (see $\textbf{Table S1}$ in the Supporting Information).

\begin{table}
		\caption{The EBRs induced by maximal symmetry Wyckoff positions in 37-SG. 194-lon.}
		\label{Table1}
    \begin{tabular*}{\textwidth}{@{\extracolsep{\fill}} ccccccc }
      \hline\hline
      EBRs & $\Gamma (0, 0, 0)$ & $M (1/2, 0, 0)$ & $K (1/3, 1/3, 0)$ & $A (0, 0, 1/2)$ & $L (1/2, 0, 1/2)$ & $H (1/3, 1/3, 1/2)$ \\
      \hline
      $(A'_1)_{2c}\uparrow G$ & $\Gamma_1^+\oplus\Gamma_4^-$ & $ M_1^+\oplus M_4^-$ & $K_5$ & $A_1$ & $L_1$ & $H_1$ \\
      $(A_g)_{6g}\uparrow G$ &  \makecell{$\Gamma_1^+\oplus\Gamma_3^+\oplus$ \\ $\Gamma_5^+\oplus\Gamma_6^+$} &  \makecell{$M_1^+\oplus M_1^- \oplus M_2^-  \oplus$ \\ $M_3^+ \oplus M_3^- \oplus M_4^-$} & $K_1\oplus K_2\oplus K_5\oplus K_6$ & $A_1\oplus A_3$ & $2L_1\oplus L_2$ & $H_1\oplus H_2\oplus H_3$\\
      \hline\hline
    \end{tabular*}
\end{table}

The second  step is identifying the real Chern insulator state based on the selected 422 OAIs. To confirm the real Chern topologies, we evaluate the $\nu_R$  for 2D $k_z=0$ and $k_z=\pi$ planes. For these two planes, the $\nu_R$ can be readily extracted from the parity eigenvalues at the time-reversal invariant momentum (TRIM) points on each plane, with~\cite{add54,add55}
\begin{equation}
(-1)^{\nu_R}=\prod_i(-1)^{\left\lfloor\left(n_{-}^{\Gamma_i} / 2\right)\right\rfloor},
\end{equation} where $\lfloor\cdots\rfloor$ is the floor function, and $n_{-}^{\Gamma_i}$ is the number of occupied bands with negative $\mathcal{P} \mathcal{T}$ eigenvalue at TRIM point ${\Gamma_i}$. If one finds that the $\nu_R$ is nontrivial ($\nu_R$ = 1) for $k_z=0$ and $k_z=\pi$ planes, which confirms the RCI nature for the carbon allotropes (see $\textbf{Figure~\ref{fig1}}$). Finally, we find 138 candidates are RCIs with nontrivial $\nu_R$ (see $\textbf{Table S1}$ in the Supporting Information).

\begin{figure}[h]
\centering
  \includegraphics[height=12cm]{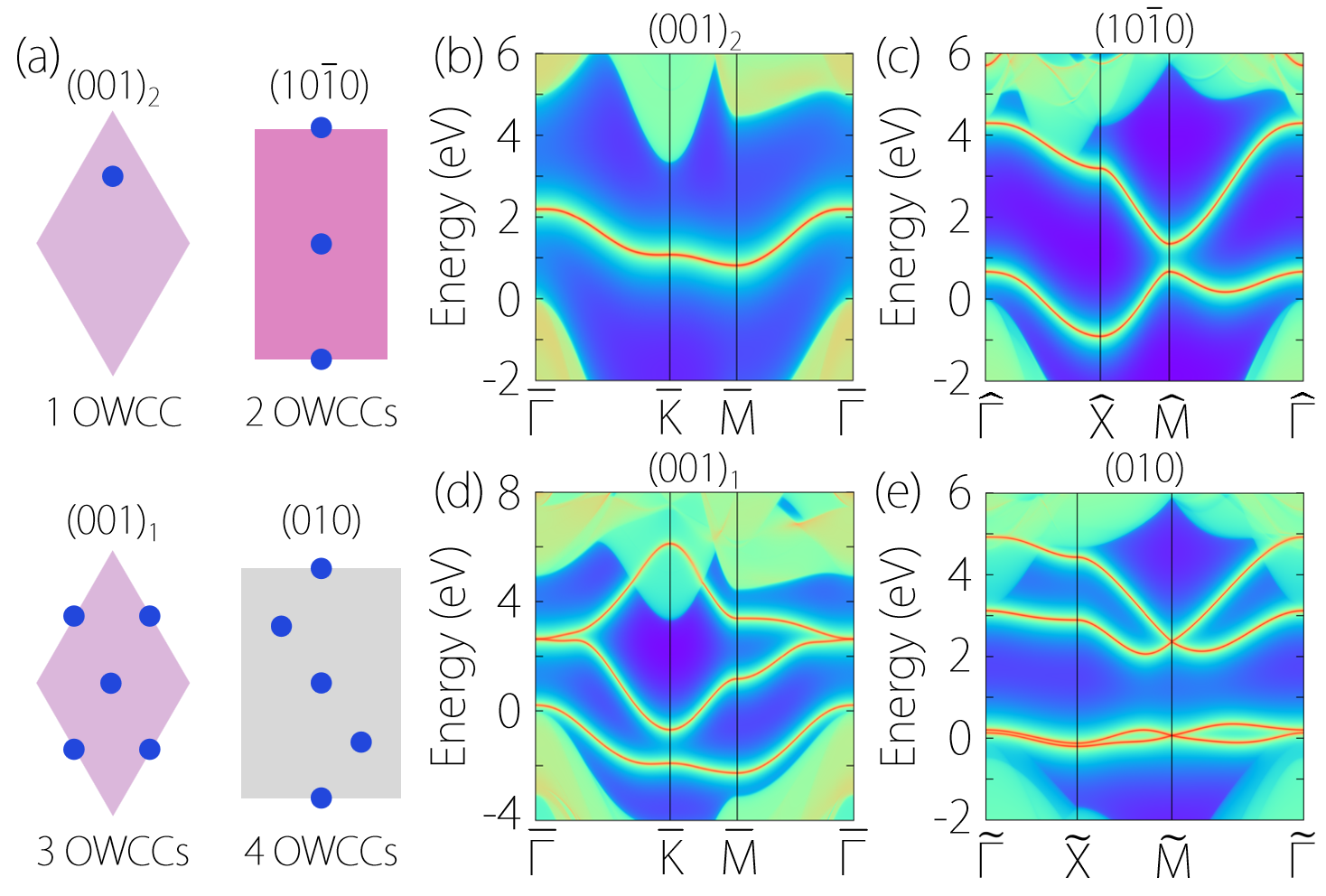}
  \caption{(a) Schematic diagrams of the numbers of OWCCs, which are cut through by (001)$_2$, $(10 \overline{1} 0)$, (001)$_1$, and (010) crystal planes. (b)-(e) 2D OSSs on (001)$_2$, $(10 \overline{1} 0)$, (001)$_1$, and (010) crystal planes, respectively. }
  \label{fig3}
\end{figure}

\begin{figure}[h]
\centering
  \includegraphics[height=11.5cm]{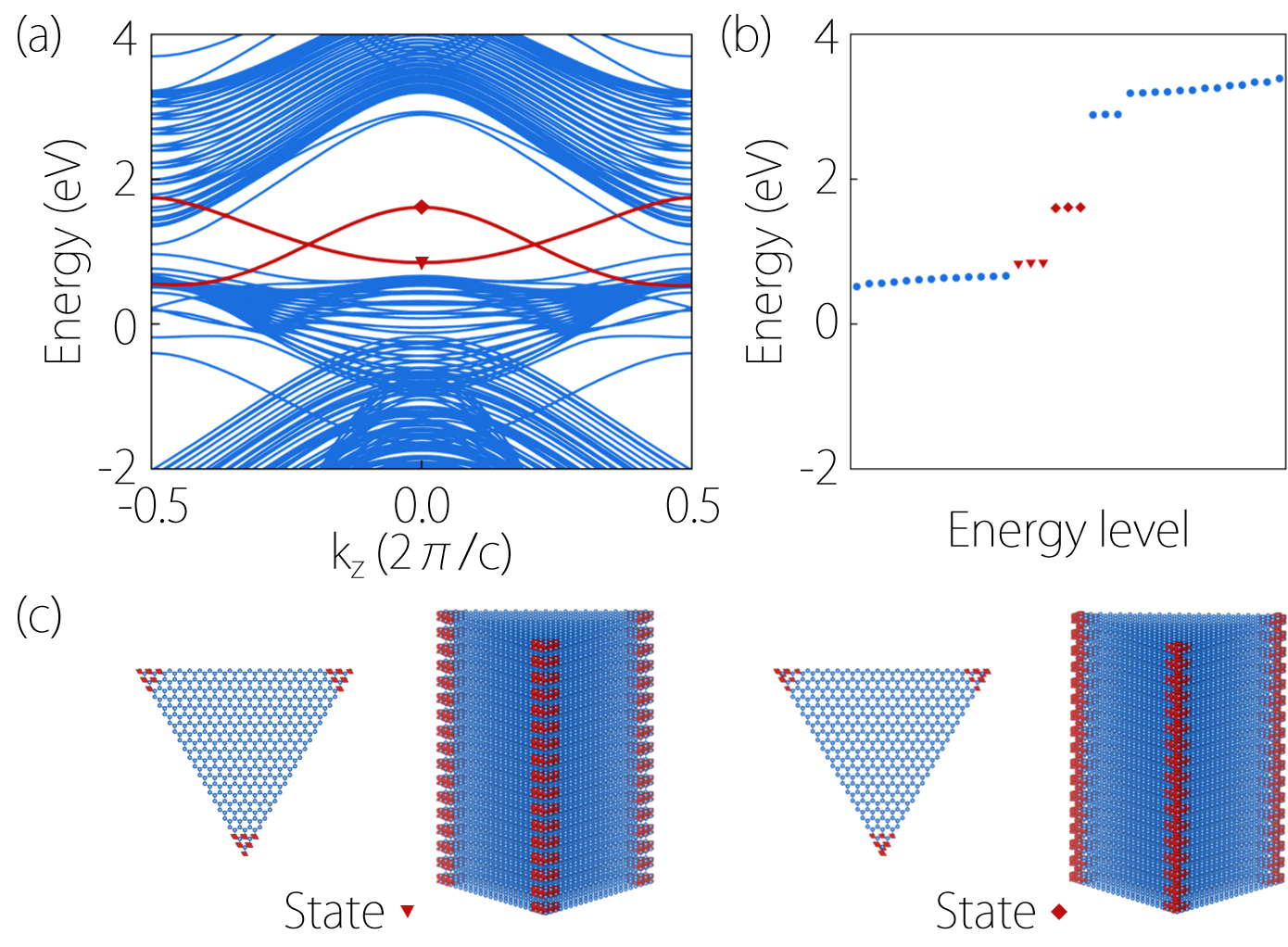}
  \caption{ (a) Calculated hinge spectrum for the 1D tube geometry of 37-SG. 194-lon sample. The hinge bands are shown in red. (b) Energy spectrum for 37-SG. 194-lon sample with a 1D tube geometry. (c) Charge spatial distributions of the two groups of hinge modes marked by triangle and quadrangle in (b) under different viewpoints.}
  \label{fig4}
\end{figure}

\subsection{3D OAIs with 2D OSSs and 1D hinges: 37-SG. 194-lon as an example}

Upon conducting an examination to calculate the BRs of the valence bands and determine the presence of OWCCs among the 703 3D carbon allotropes, it was discovered that 422 of these allotropes are 3D OAIs, as detailed in the Supporting Information (see $\textbf{Tables S2-S422}$ and $\textbf{Figures S1-S420}$). In this section, we will focus on an example, 37-SG. 194-lon, to provide a comprehensive analysis of multiple dimensions of boundaries.

The band structure is shown in $\textbf{Figure~\ref{fig2}}$c, in which the insulating state with an indirect band gap of 3.3457 eV is obvious. It is important to acknowledge that the hexagonal wurtzite-like 37-SG. 194-lon is a realistic material, initially discovered by Frondel and Marvin~\cite{add72} in the acid-insoluble residue obtained from small specimens collected from the rim of the crater. The optimized lattice constants are calculated as $a = b = 2.51$  \AA~and $c = 4.18$ \AA, which are in good agreement with the experimental data ($a = b = 2.51$  \AA~and $c = 4.12$ \AA)~\cite{add72}.

As shown in $\textbf{Table 1}$, the BRs are induced by the $A_1^{\prime}$ and $A_g$ orbitals in the D$_{6h}$ point group at Wyckoff positions 2c and 6g of SG $\mathrm{P6}_3 / \mathrm{mmc}$, which are atom-unoccupied positions in 37-SG. 194-lon. That is to say, the Wyckoff positions are 2c and 6g occupied by the OWCCs (see $\textbf{Figure~\ref{fig2}}$a). Therefore, the outcome clearly indicates that 37-SG. 194-lon is a 3D OAI.

Next, we proceed to examine the 2D boundaries, i.e., OSSs, of 37-SG. 194-lon. In contrast to the topological insulator, which exhibits topological surface states that always intersect the bulk energy gap irrespective of the system boundaries, the OSSs of the OAIs, manifest as floating bands within the bulk energy gap, exclusively emerge along the boundaries that cut through the OWCCs.	In $\textbf{Figure~\ref{fig3}}$a, we have chosen four surface planes, namely (001)$_2$, $(10 \overline{1} 0)$, (001)$_1$, and (010), which cut through 1, 2, 3, and 4 OWCCs, respectively. These surface planes are chosen to investigate the occurrence of various numbers of OSSs on the associated surface planes. The calculated OSSs on the four surface planes are shown in $\textbf{Figures~\ref{fig3}}$b-e. As expected, 1, 2, 3, and 4 OSSs appear on (001)$_2$, $(10 \overline{1} 0)$, (001)$_1$, and (010) surface planes, respectively, due to these 2D boundaries cutting through 1, 2, 3, and 4 OWCCs.

The OSSs in  $\textbf{Figures~\ref{fig3}}$b-e typically exhibit cleanliness and are situated within a large energy gap. Consequently, they can be easily separated in the energy spectrum from the bulk states, thereby facilitating theoretical research and experimental detection. The tunability of the OSS numbers can be achieved by selecting boundaries that cut through the numbers of OWCCs. This tunability is advantageous for the investigation of surface engineering. Since 2019, carbon-based catalyst~\cite{add73,add74} has been found to be an efficient, low-cost, metal-free alternative to platinum for oxygen reduction in fuel cells. Furthermore, prior studies have put forth the notion of 2D surface states~\cite{add27,add75,add76,add77,add78} as a ubiquitous catalytic descriptor in the field of topological catalysis. It is anticipated that the 422 OAIs with 2D OSSs (listed in $\textbf{Table S1}$ of the Supporting Information) may be considered as potential candidates for the future development of highly efficient carbon-based metal-free topological catalysts.

In 3D OAIs, the presence of gaps in the 3D bulk states allows for the existence of boundary states on (3-1)D surfaces. Interestingly, unconventional bulk-boundary correspondence also appears in 3D OAIs because they host second-order topological phases. That is, boundary states in two dimensions lower than the 3D bulk, i.e., (3-2)D hinges also appear in 3D OAIs. In order to determine the hinge spectrum, a tight-binding model (TB) for 1D tube geometry was built based on a sample of 37-SG. 194-lon preserving its $C_{3z}$ symmetry (see $\textbf{Figure~\ref{fig4}}$c), and then the nanotube band calculation was performed using the Wannier function~\cite{add79,add80}.  $\textbf{Figures~\ref{fig4}}$a and b illustrate the presence of two groups of threefold degenerate hinge states, denoted by red dots with triangle and quadrangle. The charge spatial distributions of these two groups of hinge states are depicted in $\textbf{Figure~\ref{fig4}}$c.
\begin{figure}[h]
\centering
  \includegraphics[height=13cm]{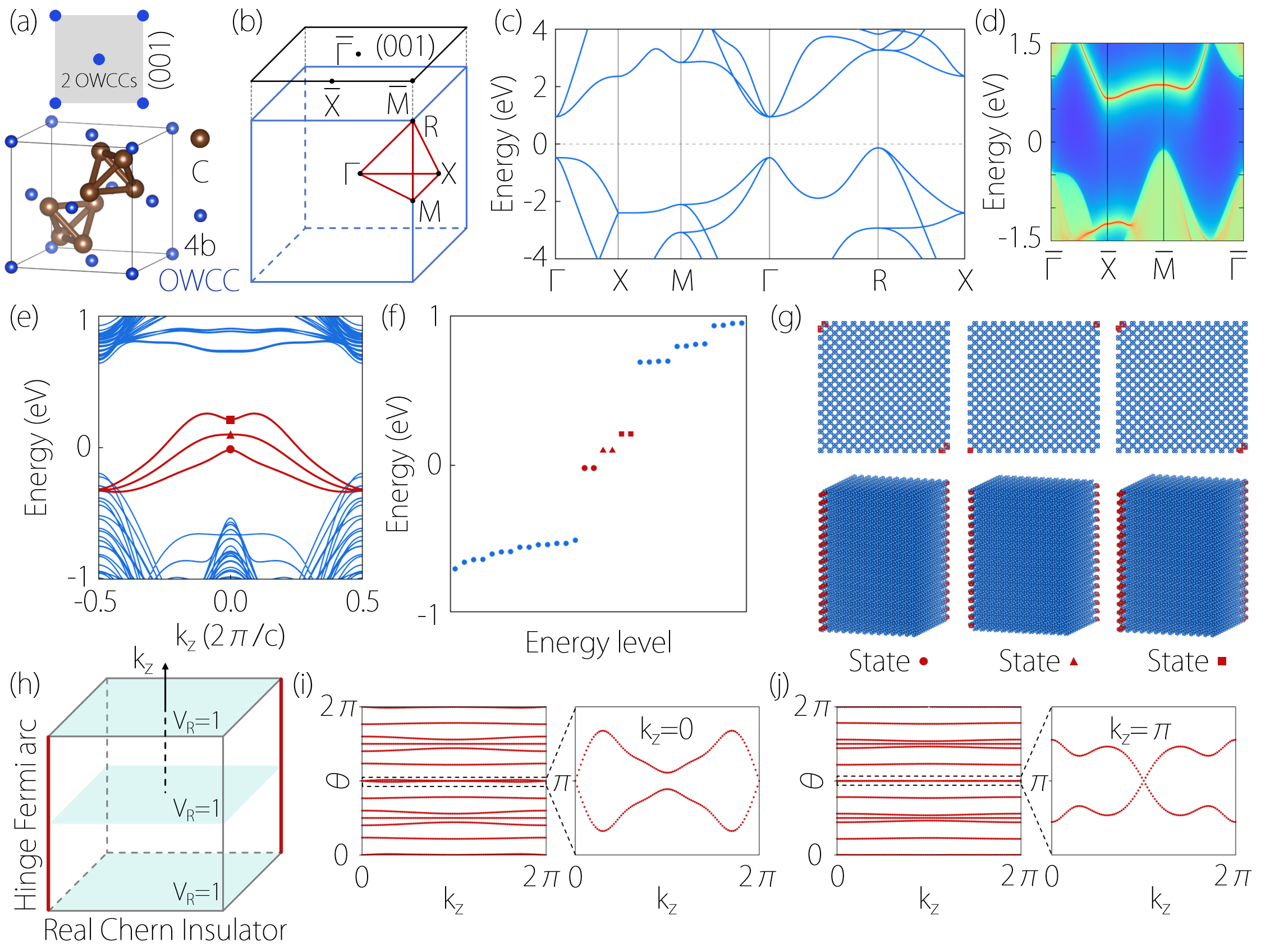}
  \caption{(a) Crystal structure of 35-SG. 224-dia-a. In (a), the OWCCs residing at the Wyckoff position 4b are denoted by small blue balls. (b) 3D bulk BZ and the projected of (001) plane. (c) Calculated band structure for 35-SG. 224-dia-a along the symmetry paths $\Gamma$--X--M--$\Gamma$--R--X. (d) 2D OSSs on (001) plane. (e) Calculated hinge spectrum for the 1D tube geometry of 35-SG. 224-dia-a sample. The three groups of hinge bands are shown in red. (f) Energy spectrum for 35-SG. 224-dia-a sample with a 1D tube geometry. (g) Charge spatial distributions of three groups of hinges marked by circle, triangle, and quadrangle in (f) under different viewpoints. (h) The schematic diagram for 3D $\mathcal{P} \mathcal{T}$-symmetric RCIs with hinge Fermi arcs, in which every 2D slice normal to $k_z$ has a nontrivial $\nu_R$. (i) and (j) Wilson loop spectrum calculated on $k_z=0$ and $k_z=\pi$ planes.
  }
  \label{fig5}
\end{figure}
3D OAIs containing 2D and 1D boundaries (surface and hinge) can be considered a useful platform for multi-dimensional boundary research. With the aid of surface-sensitive probes, such as angle-resolved photoemission spectroscopy or scanning tunneling spectroscopy, the 2D OSSs can be detected. As demonstrated in Bi~\cite{add34}, scanning tunneling spectroscopy is a useful method for examining 1D hinge modes.

\begin{table}
		\caption{Parity information of the 35-SG. 224-dia-a at the eight TRIM points.}
		\label{Table1}
\begin{tabular*}{\textwidth}{@{\extracolsep{\fill}} ccccccccc}
  \hline\hline
   & \multicolumn{4}{c}{$k_z$=0}  & \multicolumn{4}{c}{$k_z=\pi$} \\
  \cline{2-5}\cline{6-9}

  & $\Gamma$ & $X$ & $M$ & $Y$ & $Z$ & $U$ & $R$ & $T$ \\ \hline
  $n_{+}$ & 10 & 8 & 8 & 8 & 8 & 8 & 10 & 8 \\
  $n_{-}$ & 6 & 8 & 8 & 8 & 8 & 8 & 6 & 8 \\
  $\nu_R$  & \multicolumn{4}{c}{1} & \multicolumn{4}{c}{1}\\
  \hline\hline
\end{tabular*}
\end{table}

\subsection{3D RCIs with a real topology: 35-SG. 224-dia-a as an example}

 The $\mathcal{P} \mathcal{T}$ enforces a real topology for the 3D OAIs, which can be characterized by the $\nu_R$. Through a comprehensive analysis, $\nu_R$ for the $k_z=0$ and $k_z=\pi$ planes was calculated for the 422 3D OAIs. The findings revealed that among these OAIs, 138 of them displayed $\nu_R=1$, i.e., the nontrivial real topology for both planes. Further information regarding the selected 138 3D RCIs can be found in $\textbf{Tables S423-559}$ (Supporting Information).

In this section, we will focus on an example, 35-SG. 224-dia-a, to comprehensively analyze its real topology, and 2D and 1D boundaries (surfaces and hinges). 35-SG. 224-dia-a was designed by Li \textit{et al.}~\cite{add80a} by moving carbon atoms in T-carbon~\cite{add80b,add80c} half of the lattice vector along a-axis and stacking it with T-carbon, as illustrated in $\textbf{Figure~\ref{fig5}}$a. As demonstrated in $\textbf{Table S14}$ of the Supporting Information, the decomposition of the BR of dia-a into a linear combination of the EBRs in space group SG. 224 reveals the occurrence of the OWCCs (represented by blue balls in $\textbf{Figure~\ref{fig5}}$a) in 35-SG. 224-dia-a, which are located at the Wyckoff position 4b. The insulating band structure for 35-SG. 224-dia-a with an indirect bulk gap of 1.0836 eV is exhibited in $\textbf{Figure~\ref{fig5}}$c. Due to the 2D surface plane (001) cuts through two OWCCs, two OSSs can be observed, as depicted in $\textbf{Figure~\ref{fig5}}$d. Significantly, the OSSs depicted in $\textbf{Figure~\ref{fig5}}$d exhibit a gap near the Fermi level. This characteristic becomes advantageous when observing the hinge Fermi arcs in the system.  $\textbf{Figures~\ref{fig5}}$e and f illustrate the presence of three groups of twofold degenerate hinge states, denoted by red dots with circle, triangle, and quadrangle. The charge spatial distributions of the three groups of hinges are depicted in $\textbf{Figure~\ref{fig5}}$g.

In addition, the 35-SG. 224-dia-a hosts a nontrivial real topology.  In order to validate the real topology, it is convenient to perform calculations of the $\nu_R$ for the $k_z=0$ or $k_z=\pi$ plane using the parity information at the eight TRIM points. In $\textbf{Table 2}$, $n_{+}\left(n_{-}\right)$ shows the numbers of the valence states with positive (negative) $\mathcal{P} \mathcal{T}$-eigenvalues for $k_z=0$ and $k_z=\pi$ planes. Based on Eq. (3), the calculated $\nu_R$ for the $k_z=0$ and $k_z=\pi$ planes is 1, reflecting the nontrivial real topology for both planes. Actually, every 2D slice normal to $k_z$ has a nontrivial $\nu_R$, as shown in the schematic diagram for 3D RCIs in $\textbf{Figure~\ref{fig5}}$h. Furthermore, the $\nu_R$ can also be evaluated using the Wilson-loop method. We calculate the Wilson loop spectrum on $k_z=0$ and $k_z=\pi$ planes for 35-SG. 224-dia-a, as shown in $\textbf{Figures~\ref{fig5}}$i and j. Only one crossing point at $\theta$ = $\pi$ of the Wilson loop spectrum in both planes indicates $\nu_R$ = 1 for both. These results are consistent with those obtained by the parity information at the TRIM points in $\textbf{Table 2}$.

\section{Conclusion}

In summary, from first-principle calculations and theoretical analysis based on the SACADA, it has been determined that out of a total of 703 carbon allotropes, 422 are 3D OAIs. The atom-unoccupied WPs (i.e., OWCCs) in these 422 OAIs are also provided as a reference for subsequent research inquiries. We have chosen a specific example, 37-SG. 194-lon, to conduct an investigation on the relationship between the bulk and boundaries in OAIs. Moreover, the numbers of 2D OSSs can be determined by the numbers of the cut OWCCs by the 2D boundary. The tunable OSSs in these 422 OAIs can be utilized for the purpose of exploring surface engineering and developing efficient carbon-based metal-free topological catalysts in subsequent research endeavors. Moreover, the 37-SG. 194-lon hosts second-order topological phases, which induce unconventional bulk-boundary correspondence and are visible as hinge Fermi arcs. The coexistence of 2D OSSs  and 1D hinges in 37-SG. 194-lon benefits the follow-up investigations of  multi-dimensional topological boundaries. Further screening reveals that 138 of 422 OAIs host a real topology, which is characterized by a nontrivial $\nu_R$. These 138 $\mathcal{P} \mathcal{T}$-symmetric 3D RCIs are promising candidates for exploring the fascinating physics of real topological phases. Our findings show a complete list of 3D carbon-based OAIs and RCIs, and provide guidance for their use in a variety of fields, such as surface engineering, metal-free topological catalysts based on carbon allotropes, multi-dimensional boundaries, and nontrivial real band topology.

\section{Calculation methods}
We perform the first-principles calculations in the framework of density-functional theory (DFT) within the Vienna Ab initio simulation package (VASP)~\cite{add81,add82}. The exchange-correlation effect is treated in the generalized gradient approximation (GGA) in the Perdew Burke Ernzerhof (PBE) function~\cite{add83}. All calculations are performed with a plane-wave cutoff of 500 eV, and the convergence criterion for the electronic self-consistence loop was set to be 10$^{-6}$ eV, and for the structural relaxation, the Hellmanne-Feynman forces on each atom were taken to be -0.01 eV/{\AA}. The boundary bands were calculated based on the iterative Green functions method, realized using the WANNIERTOOLS package~\cite{add79}. The irreducible representations of electronic states from DFT results were calculated using the irvsp code~\cite{add84}.


\medskip
\textbf{Acknowledgements} \par 
W. Wang thanks the National Key R$\&$D Program of China (Grant no. 22022YFA1402600) for support. X. Wang thanks Australian Research Council Discovery Early Career Researcher Award (Grant no. DE240100627) for support. S. Qian thanks the National Natural Science Foundation of China (No. 12174003) and the Foundation for Distinguished Young Scholars in Higher Education Institutions of Anhui Provincial (No. 2022AH020019) for support.

\medskip

%


\end{document}